# Tunable single photon and two-photon emission in a four-level quantum dot-bimodal cavity system

Xiang Cheng, Chengwang Zhao, Han Ye, Jingran Liu, Yumin Liu, Zhongyuan Yu*

*State Key Laboratory of Information Photonics and Optical Communications, Beijing University of Posts and Telecommunications, P.O. Box 72, Beijing, 100876, P. R. China*

*E-mail: yuzhongyuan30@hotmail.com

**Abstract:** We investigate the generation of single photons and photon pairs in a cavity quantum electrodynamics system of a four-level quantum dot coupled to bimodal cavity. By tuning frequencies and intensity ratio of the driving lasers, sub-Poissonian and super-Poissonian photon statistics are obtained in each nondegenerate cavity mode respectively. Single photon emission is characterized as zero-delay second-order correlation function $g^2(0) \sim 0.15$. Photon pair emission under the two-photon resonance excitation is quantified by Mandel parameter as $Q \sim 0.04$. The mean cavity photon number in both scenarios can maintain large around 0.1. As a result, single photon emission and two-photon emission can be integrated in our proposed system only by tuning the external parameters of the driving lasers.

## 1. Introduction

Rapid developments in quantum communication and quantum information processing in recent years provide a significant incentive to develop practical non-classical light sources generating single and paired photons. A single quantum dot (QD) in a cavity has been proved to be a very promising single photon source with high brightness and indistinguishability and plays a significant role in quantum communication [1,2], quantum metrology [3] and fundamental quantum mechanics [4]. Also, quantum light sources exhibiting two-photon emission, especially entangled photon pairs, are essential building blocks for quantum information processing protocols [5], teleportation [6], cryptography or imaging [7,8]. To date, most sources of photon pairs employed are based on parametric down-conversion [9-11]. However, these sources suffer from the major drawback that the number of photon pairs generated in each process exhibits Poissonian statistics, with a non-zero probability of generating zero pair or more than one pair [12]. Promising candidate to overcome this difficulty is the QD-cavity coupled system which naturally emit photon pairs in a radiative cascade [13]. And moreover, as it is based on semiconductors, the QD-cavity coupled system has great potential for optical access, on-chip integration and scalable technological implementations [14].

Therefore, a tunable quantum light sources can be constructed by integrating the single-photon emission with two-photon emission in one device based on QD-cavity system [15]. Two-photon generation has been demonstrated by tuning the cavity frequency into resonance with half the biexciton energy [16]. Due to large biexciton binding energy, the single-photon processes are detuned and suppressed, while simultaneous two-photon emission is Purcell enhanced [17]. On the other hand, the single-photon emission of QD-cavity coupled system is mainly relied on photon blockade effect (the transition of quanta number from 1 to 2 is inhibited due to the presence of the first one), due to the strong nonlinear interaction between QD and cavity [18,19].

We propose a quantum light source scheme that integrates single-photon emission with two-photon emission by means of a four-level QD-bimodal cavity system. Thanks to the diamond type four-level energy structure of the QD, we manage to excite the single-photon and two-photon transitions separately in the same system. By studying the second-order correlation function and Mandel parameter of each cavity modes, we demonstrate the single photon and two-photon emission by tuning the ratio and frequency of the driving lasers.

## 2. Theoretical model and calculation method

The system under consideration consists of a nondegenerate bimodal microcavity containing a single four-level QD. The QD is coupled to both orthogonally polarized cavity modes *a* and *b*. We assume that the cavity modes are nondegenerate and there is no coupling between them as an ideal. Fig. 1 shows the energy level scheme of the QD-bimodal cavity coupled system. The QD states are composed of a biexciton state $|XX\rangle$, two single exciton states with orthogonal polarizations $|X\rangle$, $|Y\rangle$, and a ground state $|G\rangle$. The frequencies of two cavity modes are set to the frequency of the exciton $\omega_a = \omega_X$ and half frequency of the biexciton $\omega_b = \omega_X - \chi/2$ respectively, where $\chi$ is the biexciton binding energy. A convenient way of pumping the QD-cavity system is via a continuous excitation of the wetting layer, which will result in homogeneous broadening of the excited state and reduction of coherence time. Here, we use coherent excitation (two continuous-wave lasers with the same frequency and orthogonal polarizations are applied to excite two cavity modes respectively) in order to preserve the phase relation in the dynamics [20,21]. The driving strength for each cavity mode can be adjusted by the strength and the polarization of the lasers.

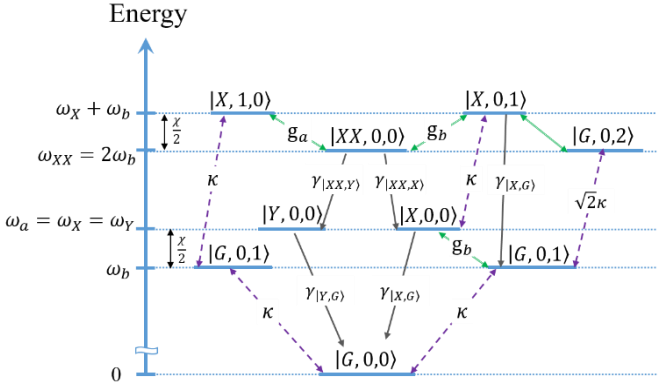

Fig. 1. Energy level scheme of a four-level QD coupled to a bimodal cavity. The microscopic configuration includes QD–cavity coupling $g_a$ and $g_b$, decay rate of cavity modes $\kappa$, excitonic spontaneous decay $\gamma$. Here, $|l,m,n\rangle$ represents the Fock state with m photons in cavity mode $a$, n photons in cavity mode $b$ and the QD at the ground state ($l = G$), exciton state ($l = X, Y$) or biexciton state ($l = XX$).

The Hamiltonian of the system under the rotating wave approximation with the laser frequency is described by ($\hbar = 1$)

$$H = H_0 + +E_a(a^\dagger + a) + E_b(b^\dagger + b), \quad (1)$$

with

$$H_0 = (2\Delta - \chi)\sigma_{XX,XX} + \Delta(\sigma_{X,X} + \sigma_{Y,Y}) + \Delta(a^\dagger a) + (\Delta - \chi/2)(b^\dagger b) + g_a(\sigma_{G,X} a^\dagger + \text{H. c.}) + g_a(\sigma_{X,XX} a^\dagger + \text{H. c.}) + g_b(\sigma_{G,Y} b^\dagger + \text{H. c.}) + g_b(\sigma_{Y,XX} b^\dagger + \text{H. c.}). \quad (2)$$

Here, $a(b)$ and $a^\dagger(b^\dagger)$ are annihilation and creation operators of the cavity mode $a$ ($b$); $\sigma_{i,j} = |i\rangle\langle j||_{i,j=G,X,Y,XX}$ is the pseudo Pauli spin operator for the QD. The exciton state of QD is assumed to be resonant to cavity mode $a$ as the detuning between the QD and cavity modes can be prevented by temperature control [22] or electrical field manipulation [23]. $\Delta$ is the detuning of the exciton state $|X\rangle$ or $|Y\rangle$ with respect to the driving lasers; $g_a(g_b)$ is the coupling strength between the QD and cavity mode $a(b)$. $E_a$ and $E_b$ are the driving laser strength for the two cavity modes, respectively. The Hamiltonian of the system states explicitly that each cavity mode only couples to photons of its corresponding mode. The dynamics of the system can be calculated using the master equation under Born-Markov approximation

$$\dot\rho = -i[H,\rho] + L\rho, \quad (3)$$

where L is the Lindblad superoperator which represents the incoherent loss of the system [24]. $L\rho$ takes the form

$$L\rho = \frac{\kappa}{2}L(a)\rho + \frac{\kappa}{2}L(b)\rho + \frac{\gamma}{2}\{L(\sigma_{G,X}) + L(\sigma_{G,Y}) + L(\sigma_{X,XX}) + L(\sigma_{Y,XX})\}\rho, \quad (4)$$

with the definition of $L(\varphi)\rho = 2\varphi\rho\varphi^\dagger - \varphi^\dagger\varphi\rho - \rho\varphi\varphi^\dagger$. Here, $\kappa$ is the decay rate for both cavity modes and $\gamma$ is the excitonic spontaneous decay between two energy states. The coupling strength are set to be equal, i.e., $g = g_a = g_b$, which can be achieved when the electromagnetic field magnitudes of both modes are equal at the location of the QD and the polarization angles between the QD diploe and both modes are equal [25]. For simplicity, the excitonic spontaneous decay $\gamma$ is set to be equal for the four-level QD.

Due to the presence of the biexciton binding energy, the frequencies of the two cavity modes are separated so that the single-photon and two-photon processes will not interfere with each other. When $\chi$ is large enough, the contributing decay processes will be reduced and only two-photon decay process survive [26]. However, at the meantime, the single-photon processes from the same system should be reserved to achieve tunable photon source. Therefore, we set $\omega_b = \omega_X - \chi/2$ to be adjusted by $\chi$ so that the two-photon resonance will maintain with various $\chi$ and cavity mode $a$ can still be in resonance with single-photon process. We adopt the experimental parameter of $\chi = 400 \mu eV$ for our following calculation, which can be achieved by a layer of InAs QDs buried at the center of a PhC double heterostructure cavity made from GaAs at the temperature of 4.5K [27].

The zero-delay second-order correlation functions $g_a^2(0) = \langle a^\dagger a^\dagger aa\rangle/\langle a^\dagger a\rangle^2$ for cavity mode $a$ and $g_b^2(0) = \langle b^\dagger b^\dagger bb\rangle/\langle b^\dagger b\rangle^2$ for cavity mode $b$ are defined to quantify the antibunching character. For cavity mode $b$, the Mandel parameter is defined as $Q = \frac{\langle n_b^2\rangle - \langle n_b\rangle^2}{\langle n_b\rangle} - 1 = \langle n_b\rangle(g_b^2(0) - 1)$, where $\langle n_b\rangle = \langle b^\dagger b\rangle$ is the mean cavity photon number (similar definition for cavity mode $a$ is $\langle n_a\rangle = \langle a^\dagger a\rangle$). Since Q changes sign with the nature of the correlations (positive for bunching, negative for antibunching), it is prefered to describe the two-photon character. The steady-state density matric $\rho$ can be obtained by numerically solving the master equation within a truncated Fock state, and the correlation function can be calculated using $\langle O\rangle = \text{Tr}(O\rho)$ which evaluates the average value of an arbitrary operator $O$ in steady state [28-30].

## 3. Results and discussions

First, we study the characteristic spectral profile of the QD-bimodal cavity system. Fig. 2 shows the normalized emission intensity of the proposed system, with different excitation scenarios corresponding to the emission spectrum peaks depicted in the inset. The emission intensity is strongly enhanced at $\Delta = 10g = \chi/2$, and two minor peaks occur at $\Delta = \pm g$. According to the energy structure of the system, $\Delta = \chi/2$ corresponds to the resonant excitation of cavity mode $b$ which gives rise to two-photon emission. The peaks at $\Delta = \pm g$ can be the scenario of the off-resonant excitation to cavity mode $a$. This is similar to the photon blockade in J-C model, which contributes to single photon transition process. It is noticeable that the peak at $\Delta = \chi/2$ has nearly twice the intensity of the peaks at $\Delta = \pm g$. This can be explained that the single photon and two-photon emission are both originated from the biexciton state, the system will emit the same amount of energy despite the different transition processes. The three peaks are spectrally narrow and isolated from each other, thus the system has an appealing practicability to be served as a tunable photon source.

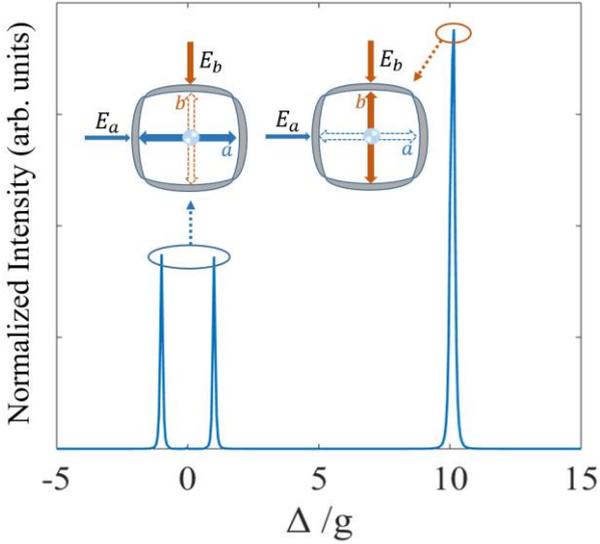

Fig. 2. Normalized emission spectrum as a function of $\Delta$. We set $\kappa = g = \chi/20 = 20\mu eV$ and $\gamma = 0.01g = 0.2\mu eV$ for the numerical calculation. Insets show the different excitation scenarios for single photon and two-photon resonance.

### 3.1. *Single photon emission*

A qualified single photon source requires sub-Poissonian photon statistics ( $g^2(0) \approx 0$ ) and large photon number emission. Thus, we can consider $g^2(0)/\langle n \rangle$ as the figure of merit for single photon source. According to the emission spectrum of the system shown in Fig.2, we can obtain single photon emission at the detuning $\Delta = g$ [31]. Fig. 3(a) plots the $g_a^2(0)$ and $g_a^2(0)/n_a$ of cavity mode $a$ at $\Delta = g$ as the functions of the driving laser intensity ratio $r = E_b/E_a$. Photons of cavity mode $a$ exhibit sub-Poissonian statistics ($g_a^2(0) < 1$), and $g_a^2(0)$ curve drops with the intensity ratio r. To explain the photon antibunching of the bimodal system, we introduce a cavity-mode structure with basis $\alpha = (a + b)/\sqrt{2}$ and $\beta = (a - b)/\sqrt{2}$ [24]. The cavity mode $a$, in the transformed basis, is equivalent to the output from two cavities: QD coupled cavity ($\alpha$) and empty cavity ($\beta$), combined on a beam splitter. Photons of cavity mode $\alpha$ shows super-Poissonian statistics due to photon-induced tunneling (the superposition of Fock states with small photon numbers and a strong presence of the vacuum state), while the empty cavity $\beta$ is a purely Poissonian coherent state [25]. As a result, the output of these two cavities combined on a beam splitter ( $a = (\alpha + \beta)/\sqrt{2}$) shows sub-Poissonian character. However, the amount of super-Poissonian light from cavity mode $\alpha$ is small because the dressed states separation in the energy ladder is large. With the increase of $E_b$, the Poissonian photons from empty cavity $\beta$ will be reduced and compensates the ratio of photons from cavity $\alpha$, thus the sub-Poissonian character of the combined output is enhanced as the $g_a^2(0)$ decreases with r.

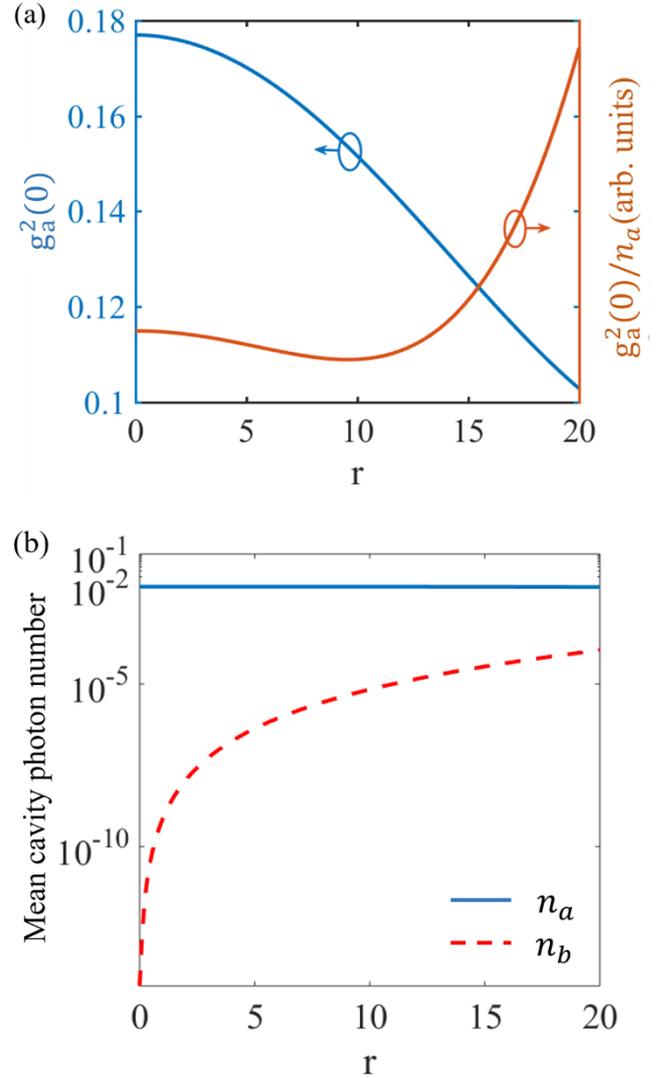

Fig.3. (a) $g_a^2(0)$ (blue solid line) and $g_a^2(0)/n_a$ (orange dash line) for cavity mode a and (b) mean cavity photon number for cavity mode a (blue solid line) and b (red dash line) as functions of driving laser intensity ratio r. $E_a$ is set to $1\mu eV$ for all the calculation.

However, as shown in Fig. 3(b), the lowest $g_a^2(0)/n$ occurs at $r = 9.5$ which is the optimal condition for photon source considering both the antibunching character and photon number. In this case, photons from cavity mode $a$ shows good photon antibunching ($g_a^2(0) \approx 0.15$) while the intracavity photon number can be achieved ~0.01. Meanwhile, the mean photon number for cavity mode $a$ is three orders of magnitude larger than that of cavity mode $b$ at $E_b/E_a = 9.5$. Even when the driving laser $E_b$ is sufficiently high ($E_b/E_a = 20$), the photon number of cavity mode $a$ is still at least two orders of magnitude larger. It demonstrates that the intracavity photon number for cavity mode $b$ is well suppressed for the single photon emission.

Since our energy schematic of the cavity is nondegenerate, we then examine the influence of frequency splitting between two cavity modes on single photon emission under the optimal excitation condition. We set $\Delta_a$ the detuning between cavity mode a and exciton state while $\Delta_b$ the detuning between cavity mode b and exciton state. The $g_a^2$ for cavity mode a in logarithmic scale as a function of $\Delta_a$ and $\Delta_b$ is illustrated in Fig. 4. Photon antibunching ($g_a^2(0) < 1$) can be achieved in blue area where $\Delta_b$ is larger than $\Delta_a$. This sub-Poissonian statistics also results from the combined output of the cavity mode $\alpha$ and $\beta$. While the frequency splitting between the two cavity modes increases, the coherent light from the empty cavity $\beta$ will have less interference with the super-Poissonian light thus leads to stronger photon antibunching. Based on the energy scheme of our system with $\omega_b = \omega_X - \chi/2$, the photon antibunching can be maintained within a range of ~0.5g for $\Delta_a$ as shown between the two red dash lines in Fig. 4. This also demonstrates the robustness of our system that a fabrication error of around 2.4GHz for the frequency of cavity mode $a$ can be tolerated.

### 3.2. Two-photon emission

Next, we investigate the two-photon emission of the system with detuning of the driving lasers and exciton state set to $\Delta = \chi/2$. As the driving lasers frequency are identical to the two-photon resonance condition $\omega_b - \omega_X = \chi/2$, the biexciton state will be excited with high probability and two-photon emission will be Purcell enhanced. Mandel parameter Q for cavity mode $b$ and mean cavity photon number for both cavity modes $a$ and $b$ are illustrated in Fig. 5. Mandel parameter Q remains 0 when driving laser intensity ratio r is below 10, then climbs to the peak (Q $\approx$ 0.04) at $r = 19.6$ and declines to negative afterwards, indicating that two-photon emission is suppressed and the photons of cavity mode $b$ exhibit sub-Poissonian statistics. Based on the transformed cavity mode basis, photons from cavity mode $b$ can be regarded as the combination of antibunched photons via QD-cavity mode coupling and superbunched photons from direct driving of mode $b$, and only the latter component can be manipulated by the ratio r. When r is small, the superbunched photons from the driving laser $E_b$ compensates the antibunched photons from the QD-cavity mode coupling and the interference between them will results in Poissonian statistics for cavity mode $b$. More superbunched photons are generated as r increases, the required compensation for the antibunched photons is already fulfilled and the emitted photons will exhibits bunching character. However, when the superbunched photons are overwhelming, the intracavity photons introduced in cavity mode $b$ will instead suppress the two-photon transition $|XX, 0,0\rangle \rightarrow |G, 0,2\rangle$ due to the population saturation of the biexciton state. Moreover, the emitted photons of cavity mode $b$ shows weak antibunching character when $r > 26$. The two-photon transition is fully suppressed ($Q < 0$) by now, the output of the cavity mode $b$ is the combination of antibunched photons from a QD-cavity mode coupling and coherent state introduced by the driving laser, which leads to weak sub-Poissonian statistics ($g_b^2(0) \approx 0.9$ at $r = 30$). The mean cavity photon number for cavity mode $b$ increases with r due to the two-photon resonance excitation of the driving laser when $r < 19.6$. Since then, the two-photon emission is suppressed but photons of cavity mode $b$ are contantly brought in by the driving laser, therefore the photon number is dominated by the driving intensity. Meanwhile, the intracavity photons of cavity mode $a$ are well suppressed as mean cavity photon number keeps around $10^{-9}$ for different driving laser intensity ratio. Hence, under the two-photon resonance excitation ($\Delta = \chi/2$), we can obtain two-photon emission from cavity mode $b$ with mean cavity photon number ~0.01 when $r = 19.6$.

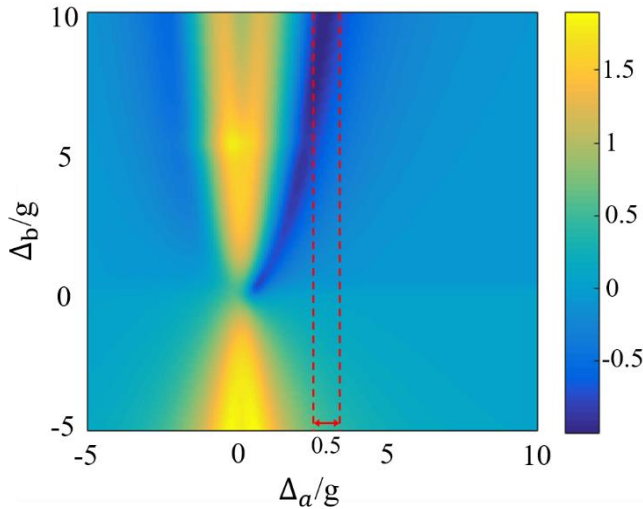

Fig. 4. $g_a^2(0)$ for cavity mode $a$ in logarithmic scale as a function of the detuning of cavity modes and exciton state.

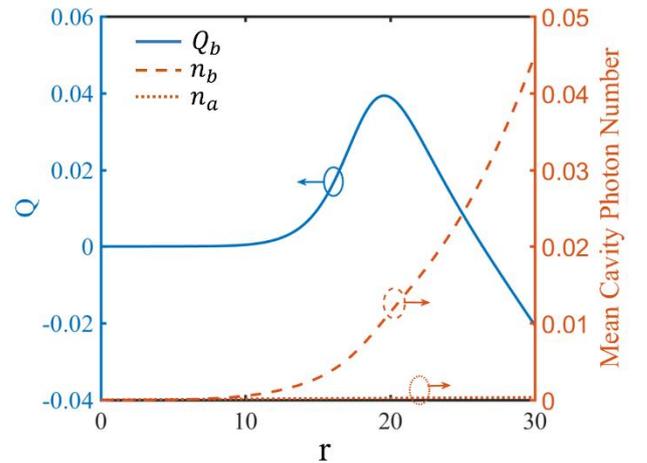

Fig. 5. Mandel parameter Q (blue solid line), mean cavity photon number for cavity mode $b$ (orange dash line) and mean cavity photon number for cavity mode $a$ (orange dot line) as functions of driving laser intensity ratio r.

## 4. Conclusion

In conclusion, we proposed a scheme to integrate single-photon emission with two-photon emission by means of a four-level QD-bimodal cavity system. By tuning the frequencies and the intensity ratio between the two driving lasers of mode $a$ and $b$, antibunched light and bunched light can be generated from each cavity mode. Single photon emission from cavity mode $a$ can be achieved when $r = 9.5$ and $\Delta = g$ as zero-delay second-order correlation function $g^2(0) \approx 0.15$ and mean cavity photon number ~0.01, while two-photon emission from cavity mode $b$ can be obtained when $r = 19.6$ and $\Delta = 10g$ as Mandel parameter $Q \approx 0.04$ and mean cavity photon number ~0.01. The emitted single photons and photon pairs can be easily distinguished by the different polarizations of each mode. Moreover, our system exhibits a good tolerance of the frequency splitting for cavity mode $a$. Thus, our proposed scheme paves a way to the applications of integrated quantum light source and quantum logic gates.

**Acknowledgments**

This work was supported by the National Natural Science Foundation of China (Grants No.61372037, No. 61671090), the Fund of State Key Laboratory of Information Photonics and Optical Communications (Beijing University of Posts and Telecommunications), P. R. China (IPOC2017ZZ04). Xiang Cheng acknowledges the funding supported by China Scholarship Council.